\documentclass{IEEEtran4PSCC}

\usepackage{tabularx}

\ifCLASSINFOpdf
   \usepackage[pdftex]{graphicx}
\else
   \usepackage[dvips]{graphicx}
\fi

\usepackage[cmex10]{amsmath}

\usepackage{xcolor}
 \usepackage[moderate,tracking=normal]{savetrees}

\usepackage{caption}
\usepackage{subcaption}
\usepackage{booktabs}
\makeatletter
\let\old@ps@headings\ps@headings
\let\old@ps@IEEEtitlepagestyle\ps@IEEEtitlepagestyle
\def\psccfooter#1{%
    \def\ps@headings{%
        \old@ps@headings%
        \def\@oddfoot{\strut\hfill#1\hfill\strut}%
        \def\@evenfoot{\strut\hfill#1\hfill\strut}%
    }%
    \def\ps@IEEEtitlepagestyle{%
        \old@ps@IEEEtitlepagestyle%
        \def\@oddfoot{\strut\hfill#1\hfill\strut}%
        \def\@evenfoot{\strut\hfill#1\hfill\strut}%
    }%
    \ps@headings%
}
\makeatother

\psccfooter{%
        \parbox{\textwidth}{\hrulefill \\ \small{23rd Power Systems Computation Conference} \hfill \begin{minipage}{0.2\textwidth}\centering \vspace*{4pt} \includegraphics[scale=0.06]{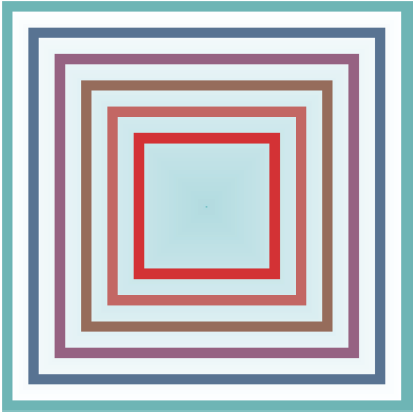}\\\small{PSCC 2024} \end{minipage} \hfill \small{Paris, France --- June 4 -- June 7, 2024}}%
}

\begin{document}

\title{Real-time Risk Prediction of Cascading Blackouts with Graph Neural Networks}
\title{Cascading Blackout Severity Prediction with Statistically-Augmented Graph Neural Networks}

\author{\IEEEauthorblockN{Joe Gorka\IEEEauthorrefmark{1}\IEEEauthorrefmark{3},
Tim Hsu\IEEEauthorrefmark{2},
Wenting Li\IEEEauthorrefmark{3}, 
Yury Maximov\IEEEauthorrefmark{3}, and
Line Roald\IEEEauthorrefmark{1}}


\IEEEauthorblockA{\IEEEauthorrefmark{1} University of Wisconsin, Madison,USA
}

\IEEEauthorblockA{\IEEEauthorrefmark{2} Lawrence Livermore National Laboratory
}

\IEEEauthorblockA{\IEEEauthorrefmark{3} Los Alamos National Laboratory
}

}


\maketitle



\thanksto{\noindent The work is supported by the Advanced Grid Modeling Program of the U.S. Department of Energy’s Office of Electricity as part of the “Robust Real-Time Control, Monitoring, and Protection of Large-Scale Power Grids in Response to Extreme Events” project.}

\vspace{-10pt}

\begin{abstract}
Higher variability in grid conditions, resulting from growing renewable penetration and increased incidence of extreme weather events, has increased the difficulty of screening for scenarios that may lead to catastrophic cascading failures. Traditional power-flow-based tools for assessing cascading blackout risk are too slow to properly explore the space of possible failures and load/generation patterns. We add to the growing literature of faster graph-neural-network (GNN)-based techniques, developing two novel techniques for the estimation of blackout magnitude from initial grid conditions. First we propose several methods for employing an initial classification step to filter out safe `non-blackout' scenarios prior to magnitude estimation. Second, using insights from the statistical properties of cascading blackouts, we propose a method for facilitating non-local message-passing in our GNN models. We validate these two approaches on a large simulated dataset, and show the potential of both to increase blackout size estimation performance. 

\end{abstract}
\section{Introduction}

As the grid experiences increasing pressure from climate-change-driven extreme weather events and greater renewable penetration, ensuring grid security has become more challenging. While operators generally require that the power system is resilient to failures of a single component, security under multiple failures or under unexpected shifts in generation or loading patterns is typically not ensured. In the worst case, an initial set of failures can lead to a ``cascading blackout'', in which a chain of subsequent failures results in significant load-shedding. 

Though several high-quality engineering simulation tools, such as OPA \cite{OPA}, DCSIMSEP \cite{Eppstein_Hines_2012}, and the Manchester Model \cite{Kirschen_2004}, can provide operators with an estimate of potential blackout size given the system state and a set of initial failures, such simulations are too slow to efficiently cover the massive space over which it is often necessary to search. To illustrate: a system with only 100 lines has 166,750 potential failures of 3 or fewer lines. Incorporating the wide range of operating states that must be considered in the presence of unpredictable renewable generation, the number of potential scenarios grows even higher. This complexity creates challenges in using such simulation-based tools to assess a system's risk of cascading blackout, particularly if such a tool is to be used for `online' security assessment. 

\subsection{Related work}
Cascading blackout simulators such as OPA and DCSIMSEP involve repeatedly solving power flow problems, the computational burden of which makes it impossible to efficiently search the space of potential system states/initiating contingencies. A wide variety of methods have been proposed to mitigate this issue. One category of techniques has employed intelligent sampling to reduce the number of flow-based simulations which must be performed. This includes approaches such as the importance sampling method \cite{Chen_Mili_2013,Henneaux_Labeau_2014}, splitting method \cite{Kim_Bucklew_Dobson_2013,Wang_Chen_Liu_Chen_Shortle_Wu_2015}, and the random chemistry method \cite{Eppstein_Hines_2012,Rezaei_Hines_Eppstein_2015}.

In contrast, others have sought to replace flow-based simulations altogether, thus speeding up the risk evaluation of each individual scenario. Such approaches generally fall into the categories of heuristic, statistical, and machine-learning methods. Heuristic methods include earlier works such as \cite{Dobson2003} and \cite{Dobson_Carreras_Newman_2004}, in which it is assumed that lost load is distributed evenly over all components after a failure, and more recent `percolation' approaches which assume local propagation of failures \cite{Xiao_Hongda_Yeh_2011,Kong_Yeh_2012,Wang_Zhou_Hu_2018,Zhang_Yeh_Modiano_2019}. While the use of heuristic models can yield a large speed-up, they do not capture the physics or control actions that define the behavior of failure states in real power systems.

Statistical methods for blackout size estimation, on the other hand, make use of large quantities of data generated by power-flow-based engineering simulations to gain insight into the qualities and dynamics of cascading blackouts. Such methods include the so-called `Influence' model \cite{hines_cascading_2017,Zhou_Kai_Dobson_2020}, and `Interaction' model \cite{Qi_Junjian_2015,Ju_Qi_Junjian_2015}, in which conditional statistical models are employed to predict the next step of a cascade given currently-failed components. As they are based on data from engineering simulations, statistical methods may better capture the physics/control actions present in the system. They have also led to some interesting insights, such as the observation of significant non-local failure propagation \cite{hines_cascading_2017}, casting further doubt on the validity of `percolation', i.e. local contagion, models.

However, while statistical models may reveal large-scale patterns in the way that cascading blackouts tend to occur in a power system, they generally only model failure propagation conditional on a current set of component failures. This disallows the consideration of other aspects of system status, such as loading and generation, which may play a major role in determining whether an initial failure spirals into a catastrophic cascading blackout. To remedy this issue, many machine-learning (ML)-based approaches have been proposed which take as input a more complete picture of system status. Some works have used ML to speed up/augment flow-based simulators. For example, \cite{Liu_Zhang_Wu_Botterud_Yao_Kang_2021} pairs a graph neural network (GNN) with a physics-based heuristic to guide a simulator as it proceeds down the search tree of potential outage trajectories. In \cite{Zhu_Zhou_Wei_Wang_2023}, a GNN is instead trained to act as a proxy for the simulator power-flow calculation.

Purely ML-based techniques for assessing cascading blackout risk have also been suggested \cite{Shuvro_Das_Hayat_Talukder_2019,Varbella_Gjorgiev_Sansavini_2023,Zhu_Zhou_Wei_Zhang_2022,Ahmad_Papadopoulos_2022}. In \cite{Shuvro_Das_Hayat_Talukder_2019} simple ML models such as Decision Trees and Support Vector Machines are tested both for classification (blackout/no-blackout) and regression (magnitude of blackout). The best-performing models in this study show low accuracy in regression, but solid classification performance (F1 score of 0.91).

The majority of more recent ML-based works in this area have made use of GNNs, which enable the physical topology of the power system to be encoded explicitly in the input data. The advantage of GNNs over other model types, such as fully connected neural networks, is demonstrated in \cite{Varbella_Gjorgiev_Sansavini_2023}, showing computational advantages as well as a much greater ability to generalize to multiple systems and varying grid topologies. In \cite{Zhu_Zhou_Wei_Zhang_2022} a GNN is combined with a long-short-term-memory (LSTM) model to perform both classification (blackout/no blackout) and regression to estimate VaR and cVar. A similar GNN+LSTM approach is taken in \cite{Varbella_Gjorgiev_Sansavini_2023} to perform classification on a variety of different test systems and initial outage sizes. Finally, in \cite{Ahmad_Papadopoulos_2022}, a GNN+LSTM model is combined with a trainable `importance mask', allowing the model to learn which of the edges in the physical graph are most
important for facilitating information transfer relevant to the
target variables (in this case the classification of yes/no blackout).

\subsection{Contributions:}
First, motivated by the fact that an overwhelming majority of failures encountered in normal operation of a power system do not lead to significant blackout \cite{Chen_Thorp_Parashar_2001,Carreras_Newman_Dobson_Poole_2001,Talukdar_Apt_Ilic_Lave_Morgan_2003}, we investigate the impact of such an unbalanced distribution (heavily favoring 'no blackout' samples) on the task of learning to estimate blackout risk. Specifically, we work to determine whether an initial classification step may be employed to increase the efficacy of blackout magnitude estimation. We propose three novel GNN-based models. The first consists of a single GNN which is trained on both blackout and non-blackout samples, and provides blackout estimates in a single step. The second employs an initial classification step, using a GNN trained only on blackout samples to estimate the blackout magnitude of samples classified as `positive'. And the third adds a `verification' process, making use of two separate GNN models, to avoid problems arising from classifier false negatives. Noting the results of \cite{Shuvro_Das_Hayat_Talukder_2019}, we perform classification using only a simple non-linear classifier (XGBoost).

Secondly, inspired by the `importance-masking' of \cite{Ahmad_Papadopoulos_2022} and the non-local failure relationships described in \cite{hines_cascading_2017}, we propose and test a new method for moving beyond the physical grid topology when designing the input graph of our GNN models. Specifically, we propose the augmentation of this physical topology with `statistical edges' which seek to facilitate information flow between non-local (but statistically associated) areas of the grid.

Our work is distinguished from the already existing literature in the following ways:
\begin{enumerate}
    \item To our knowledge, we are the first to investigate the efficacy of a pre-regression classification step for estimation of blackout size from initial grid conditions/failures.
 
    \item To our knowledge, we are the first to employ a GNN-based model for blackout size estimation which takes into account the insight that non-local parts of the computational graph may contain mutually-relevant information. To do this we propose a novel method of augmenting the physical network topology with edges that represent statistical relationships between component failures. 

    \item Finally, we make the cascading failure dataset generated in this work publicly available at \cite{MindsUW}.

\end{enumerate}

The rest of this paper is organized as follows. Section \ref{sec:pred_BO_size} more thoroughly characterizes the problem of predicting blackout size from initial grid conditions/failures, and further motivates our investigations into pre-regression classification and GNN input-topology augmentation. Section \ref{sec:modeling_methodology} describes the design of the GNN regressor and XGBoost classifier components that make up our proposed models. Section \ref{sec:stat_topology_aug} describes our method for statistical GNN input-topology augmentation. Section \ref{sec:model_variants} details our proposed models, and Section \ref{sec:case_study} describes the case study set up, including data generation and training of the models. Section \ref{sec:results} details the results. Lastly, we give some final conclusions in Section \ref{sec:conclusions}.

\section{Predicting Blackout Size}
\label{sec:pred_BO_size}
Our goal is to predict the size of a potential blackout (if any) given only the initial state of a power system (loading, generation, and line parameters) and a set of initiating line failures. We consider only line failures, as they are an order of magnitude more likely than node outages \cite{hines_cascading_2017}. 

In this section we discuss the two main questions we are concerned with answering.

\subsection{Regression, Classification, Both?} 
The prediction of potential blackout severity is intrinsically a regression problem: we need to know the potential magnitude of an outage in order to calibrate what kind of action we are willing to take in order to prevent it. In the context of (supervised) machine-learning methods, this means that during training we expose a model to a samples with a variety of system states/sets of initiating failures, each of which has a corresponding continuous label indicating the amount of load loss (i.e. blackout size). However, if the dataset centers around typical operating states and realistic (mostly small) initial failures, it is likely to mostly contain samples that result in little to no blackout at all \cite{Chen_Thorp_Parashar_2001,Carreras_Newman_Dobson_Poole_2001,Talukdar_Apt_Ilic_Lave_Morgan_2003}. This leads to an interesting problem. We would like our model to be trained on a distribution that is as similar as possible to that of the real world, yet also want it to successfully estimate the magnitude of a small but very impactful subset of samples (those in which there is a significant blackout). 

The first focus of this work is to investigate the impact of this class imbalance (in favor of non-blackout samples) on the ability of a modern graph neural network (GNN)-based machine-learning model to successfully estimate blackout magnitude from initial grid conditions. To do this, we first assume an ability to perfectly classify samples into no-blackout and blackout categories, using the blackout samples to train a GNN. We then propose several practically-implementable GNN-based models which make use of a real classifier in place of the perfect assumption. We then compare the performance of these models to that of a GNN-based model trained on both blackout and non-blackout samples, as well as a combination of both types of models.

\subsection{Which GNN Input Topology?}
GNNs which employ spatial or localized spectral convolutions in their message-passing step, such as the GCN \cite{kipf_welling_2017}, GAT \cite{velickovic_2018}, or ChebNet \cite{defferrard_2016} models calculate embeddings (processed feature-values) of each node/edge as a function of both their previous-layer values and the previous-layer values of neighboring nodes/edges. In this way, such models encode an implicit bias that the neighbors of each component contain information relevant to the desired output of that component. This can create a problem in the case that it is desirable that information be passed between nodes/edges that are distant from each other on the graph. While the stacking of multiple GNN layers can allow information to propagate further across the graph (a GNN with $n$ layers will contain embeddings that are a function of their respective $n$-hop neighborhoods), it can also lead to the problem of so-called ``over-smoothing'' \cite{xu_hu_2019}. Over-smoothing occurs when component embeddings have incorporated so much information from elsewhere in the graph that they become essentially homogeneous, resulting in catastrophic loss of model expressivity. 

Specific to the problem of modeling cascading blackouts, reference \cite{hines_cascading_2017} titled \emph{``Cascading Power Outages Propagate Locally in an Influence Graph that is not the Actual Grid Topology"} tells us exactly that -- i.e. that cascading blackouts do not propagate between neighbouring grid nodes, but rather in a so-called influence graph. 
This presents a problem for capturing such events using graph-based models based only on the physical power system topology. While in this work we are only concerned with determining the eventual load loss outcome of an initial set of component failures, and do not explicitly model the intervening steps, 
\cite{hines_cascading_2017}
suggests that the state of grid components which are physically/graphically distant from each other may contain information relevant to determining their respective failure states.

The second focus of this work is therefore to test the impact of facilitating increased connectivity between areas of the power network that are statistically associated but not physically close enough (in terms of geodesic distance) to effectively exchange information during the message-passing step of a GNN-based model. To do this, we propose a method for augmenting the physical topology of the power system with `statistical edges'. We test the impact of this method across all GNN models mentioned in the previous section (and described in more detail in Sections III and V).

\section{Modeling Methodology}
\label{sec:modeling_methodology}
In order to answer the two questions posed in the previous section, regarding the impacts of GNN input topology and pre-regression classification step, we make use of a wide array of model variants (described in Section V). However, though employed in different ways, each model variant makes use of the same regression and classification building blocks. This section describes the GNN architecture used to estimate blackout size across all model variants, and the simple XGBoost classifier employed in those variants that make use of a pre-regression classification (blackout/no-blackout) step. 

As they are likely to be highly case-dependent, many GNN and XGBoost hyper-parameters (e.g. ) are left unspecified in this section. Sections \ref{sec:tune_GNN} and \ref{sec:clf_tune} detail the optimal values for these hyper-parameters with respect to our specific case-study, in addition to the process we followed to obtain them.

\subsection{GNN Regressor Architecture}
We use a GNN input topology that is similar to that of the power system. This topology is an un-directed graph $G(V,E)$, in which $v_i \in V$ represent the set of vertices (nodes), and $e_{i,j} \in E$ represent the set of edges. Equivalently we use $v$ and $e$ to represent, respectively, the feature values at a given node or edge. The specific node and edge features used in our case study implementation are described in Section \ref{sec:dataset}.

To estimate the magnitude of blackout risk, we make use of a simple message-passing GNN which takes as input graph-valued samples describing the state of the power system and returns a single floating-point value. Here we describe the steps of a forward pass through this architecture, noting that every described transformation is trainable unless otherwise specified.

First, all node features $v_i\in V$ and edge features $e_{i,j}\in E$ pass through an initial embedding layer in which they are transformed to 128-dimensional representations $v_i'$ and $e_i'$. Mathematically, this can be written:
\begin{equation}
    v_i' \longleftarrow f_{init,V}(v_i)
\end{equation}
\begin{equation}
    e_{i,j}' \longleftarrow f_{init,E}(e_{i,j})
\end{equation}
where $f_{init,V}$ and $f_{init,E}$ are separate 2-layer feed-forward neural networks. 

Next, the sample passes through a number of graph-convolutional message-passing layers. For these, we make use of the MeshGraphNet \cite{pfaff_2021} architecture, described as follows. At each layer, two steps are taken. First, an aggregation step is performed in which each component sums its own feature value with those of its neighbors. In the case of a node $v_i$, these neighbors are the edges $e_{i,j}$ that are connected to $v_i$. The neighbors of an edge $e_{i,j}$ are considered to be the nodes $v_i$ and $v_j$ which define the endpoints of $e_{i,j}$. Following this summation is a transformation step in which the aggregated states at each node and edge are passed through single-layer feed-forward neural networks $f_V$ and $f_E$, generating the node and edge embeddings to be taken as input to the following layer. Together, this process can be written as follows, for node and edge embeddings respectively:
\begin{equation}
    v'_i \longleftarrow f_V(v_i,\sum_j{e_{i,j}})+v_i
    \label{eqn:node_embedding_update}
\end{equation}
\begin{equation}
    e'_{i,j} \longleftarrow f_E(e_{i,j},v_i,v_j) + e_{i,j}
    \label{eqn:edge_embedding_update}
\end{equation}
Where $f_V$ and $f_E$ are single-layer feed-forward neural networks with a residual connection (indicated by the addition of the initial value in both cases).

At the conclusion of the message-passing layers, each node and edge is associated with a 128-dimensional feature-vector which together (ideally) contain information that is relevant to the estimation of our target value. It is then necessary to map this set of vectors to a single value $\hat{y}$, which will be our estimate of the magnitude of blackout resulting from the initial grid conditions/set of failures. This is accomplished via a node-pooling layer, followed by a final feed-forward neural-network output layer. Pooling is performed over nodes only, as the method of updating node-embeddings at each message-passing layer (Eqn \ref{eqn:node_embedding_update}), ensures that edge information has already been incorporated.
The pooling and final output layers can be written together in a single expression as follows:
\begin{equation}
    \hat{y} = f_{final,V}(\sum_i{v_i})
\end{equation}
In this equation, the node pooling layer is represented as a sum over the node embeddings $v_i$ as they exist following the final message-passing layer. The function $f_{final,V}$ indicates a single-layer feed-forward neural network that maps the pooled node embedding to the final output $\hat{y}$.

\subsection{Classification Model}
As our classification model, we choose to employ XGBoost \cite{chen_2016} due to its excellent track record in a wide variety of classification tasks involving data that can be expressed in a tabular format \cite{shwartz_armon_2022,Grinsztajn_2022}. Additionally, it requires little time and computational power to train as compared to neural-network-based models. 

The features used for classification in this work are the same as the GNN features, but are represented as a 1-dimensional vector without any reference to the power system topology. This is further described in Section \ref{sec:dataset}.

\section{Statistical Topology Augmentation}
\label{sec:stat_topology_aug}
\begin{figure}
    \centering
    \includegraphics[width=0.7\columnwidth]{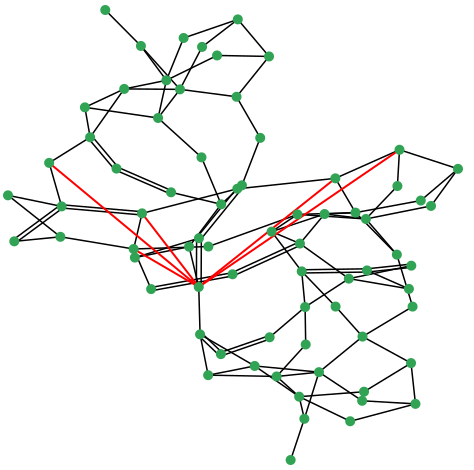}
    \caption{\small Augmented topology for the RTS-GMLC Test System. Green Buses and Black Lines represent physical power system topology, while red lines represent additional statistical lines.}
    \label{fig:rts-statlines}
\end{figure}

Our goal is to ensure that the message-passing mechanism described in the previous section yields node and edge embeddings that incorporate maximally-relevant information from elsewhere in the graph. As mentioned in Section IIA, statistical methods such as the `Influence Model' approach from \cite{hines_cascading_2017} suggest that strong associations exist between component failures that are distant along the physical network topology. For that reason we propose a method of augmenting this physical topology with the aim of facilitating more effective message-passing between such associated components. Here we focus on statistical associations between line failures, but there is no barrier to considering node failures instead. Inspired by \cite{hines_cascading_2017}, our method for augmenting the GNN topology proceeds as follows:

\begin{enumerate}
    \item Generate data: We generate a dataset of cascading failure simulation results that includes a list of all lines that failed over the course of the simulation.
    More details about specific data set generation used in our implementation is described in Sections \ref{sec:dataset} and \ref{sec:data-stat}.
    \item Calculate co-failure frequencies: From this dataset, we calculate the pair-wise frequency of line failures. This tells us which pairs of lines tend to fail together.
    \item Filter neighboring line-pairs: We filter the list of line-pairs to remove any which already share a node (and should therefore have no difficulty sharing information).
    \item Weigh co-failure frequencies by shortest-path distance: As the goal is to facilitate message-passing between areas of the graph which are both relevant to each other and topologically distant, we weight the remaining line-pair co-failure frequencies by their shortest-path distances along the physical network topology.
    \item Select final line-pairs: We select the $k$ line pairs that score the highest on our distance-weighted co-failure rate metric. The optimal value of $k$ will depend on the data.
    \item Convert final line-pairs to statistical/influence edges: For each line-pair, we select one endpoint-node from each line and add the connection between them as an edge in our graph.
\end{enumerate}
As an example of a system topology which has been augmented with statistical edges, see Fig. \ref{fig:rts-statlines}. 

\section{Model Variants}
\begin{figure}
    \centering
    \begin{subfigure}[t]{\columnwidth}
        \centering
        \includegraphics[width=0.65\columnwidth]{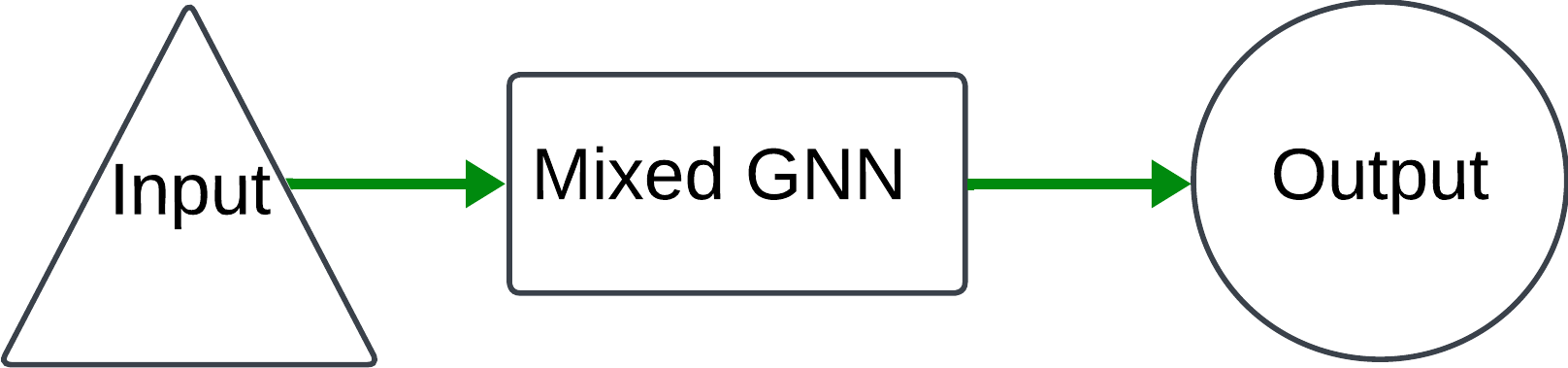} 
        \subcaption{R Model Diagram}
        \label{fig:r_diagram}
    \end{subfigure}

    \begin{subfigure}[t]{\columnwidth}
        \centering 
        \includegraphics[width=0.9\columnwidth]{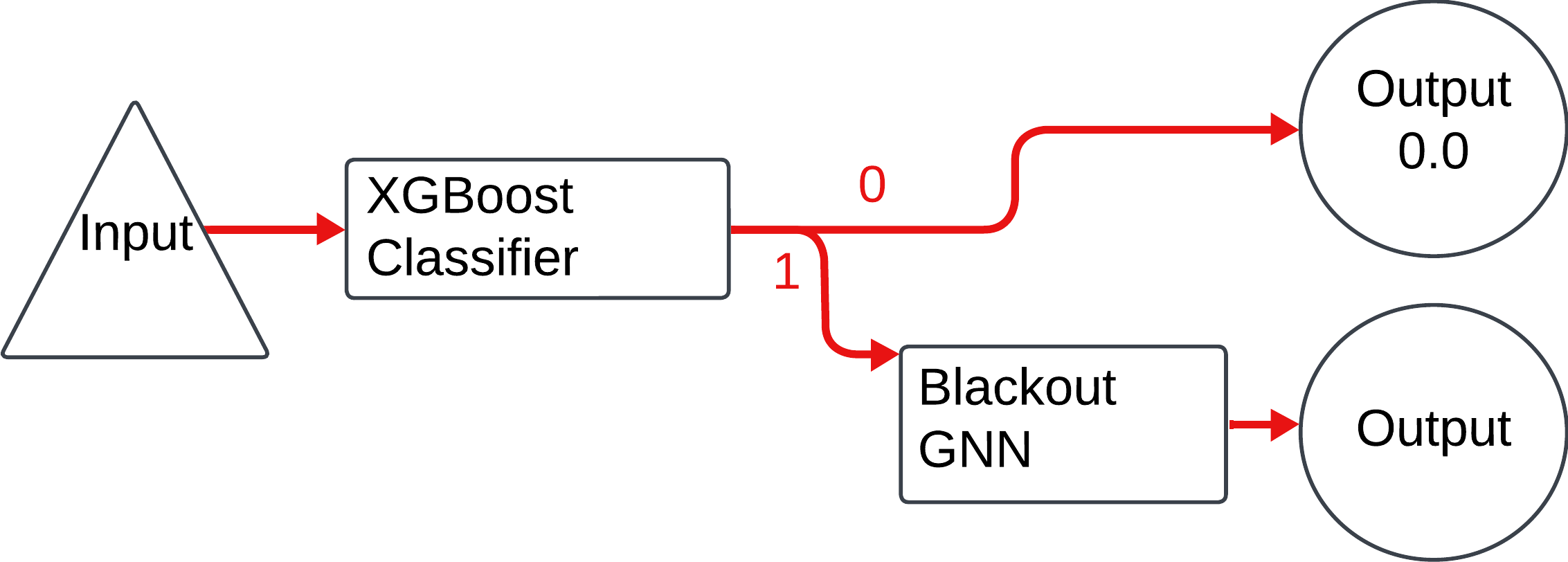}
        \caption{CR Model Diagram}
        \label{fig:cr_diagram}
    \end{subfigure}

        \begin{subfigure}[t]{\columnwidth}
        \centering 
        \includegraphics[width=0.95\columnwidth]{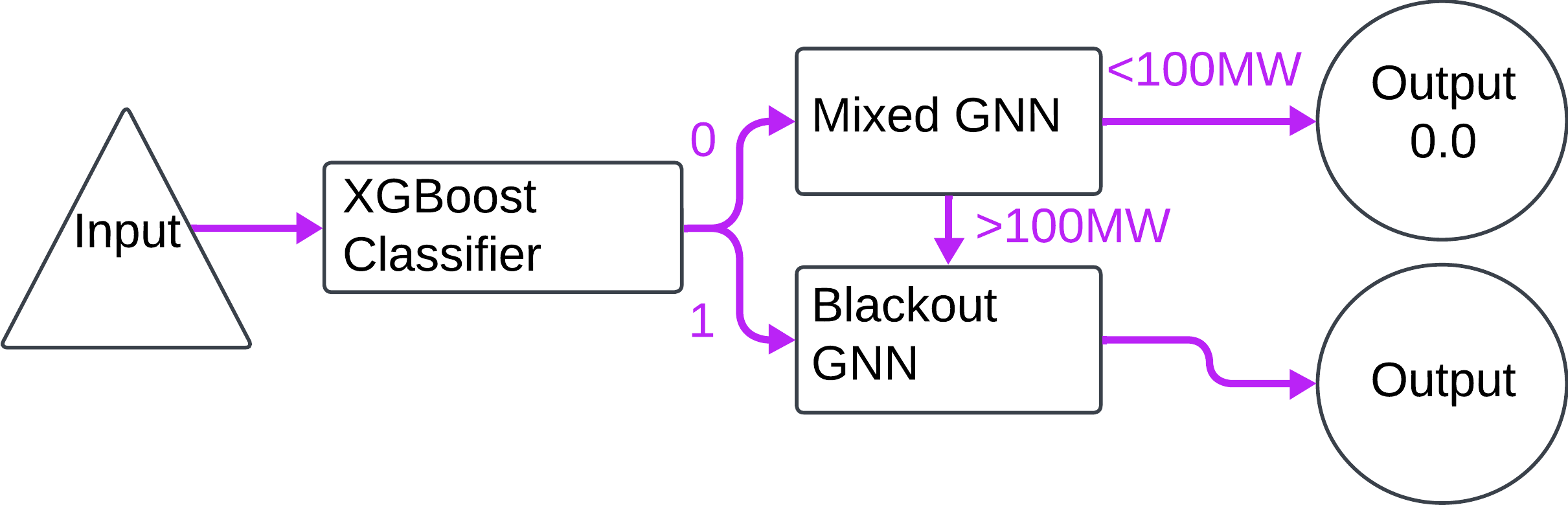}
        \caption{CVR Model Diagram}
        \label{fig:cvr_diagram}
    \end{subfigure}

\caption{R, CR, and CVR Model Diagrams}
\label{fig:model_diagrams}

\end{figure}

\label{sec:model_variants}
In this section we propose six different models based on the building blocks described above. The models are divided into three variants (denoted by R, CR, and CVR) based on if/how they employ a classification step prior to estimating blackout size. We also introduce the R+, CR+, and CVR+ variants which employ an augmented graph topology.

A visual representation of the R, CR, and CVR models can be found in Figure \ref{fig:model_diagrams}. Augmented-topology versions (marked with +) follow the same structure, differing only in the graph topology used in the GNN regressor components.
Note that in this and later sections we will refer to the GNN components of the various models as either 'mixed' (trained on both blackout and non-blackout samples), or 'blackout-only' (trained only on blackout samples.)

\subsection{Regress (R)}
In this method, a GNN model as described in Section III is trained on mixed samples, including samples in which there is a blackout, and those where there is not. For unseen samples, blackout size is estimated in a single step by passing them through this 'mixed' GNN. For the R variant, we use a graph topology which contains only physical edges (representing the presence of a physical power line).

\subsection{Regress on Augmented Topology (R+)}
As we do in the R variant above, we train a GNN to directly provide an estimate of potential blackout magnitude. However, we make use of an augmented graph topology which contains both physical and statistical edges.

\subsection{Classify then Regress (CR)}
In this variant we separate the tasks of determining whether or not there will be a blackout from the task of determining how large such a blackout may be. A classification model is trained to distinguish between blackout and no-blackout samples. Separately, a blackout-only GNN model is trained only on samples for which there is a blackout and is used to determine the blackout magnitude of a sample if it is classified 'positive' in the first step. This GNN makes use of the graph topology containing only physical edges.

\subsection{Classify then Regress on Augmented Topology (CR+)}
This variant is identical to the CR model defined above, but instead employs the augmented graph topology containing both physical and statistical edges.

\subsection{Classify, Verify, then Regress (CVR)}
The CVR model is identical to the CR model in its handling of samples classified as positive (i.e. leading to blackout). These samples are passed to a blackout-only GNN for final magnitude estimation. However in CVR, samples classified as negative (no blackout) are subject to an additional a verification step. Instead of immediately returning zero, samples classified as negative are also evaluated by a mixed GNN. If the output of this mixed GNN is high ($>$100MW), we consider the classifier output to be a probable false-negative and return the blackout magnitude estimate resulting from feeding that sample through the blackout-only GNN. We use a mixed GNN for verification because it is exposed to both blackout and no-blackout samples during training.

\subsection{Classify, Verify, then Regress on Augmented Topology (CVR+)}
Finally, we propose the augmented-topology version of the CVR model defined above.

\section{Case Study}
\label{sec:case_study}
 
\subsection{Dataset Generation}
\label{sec:dataset}

\begin{figure}
    \centering
    \includegraphics[width=0.9\linewidth]{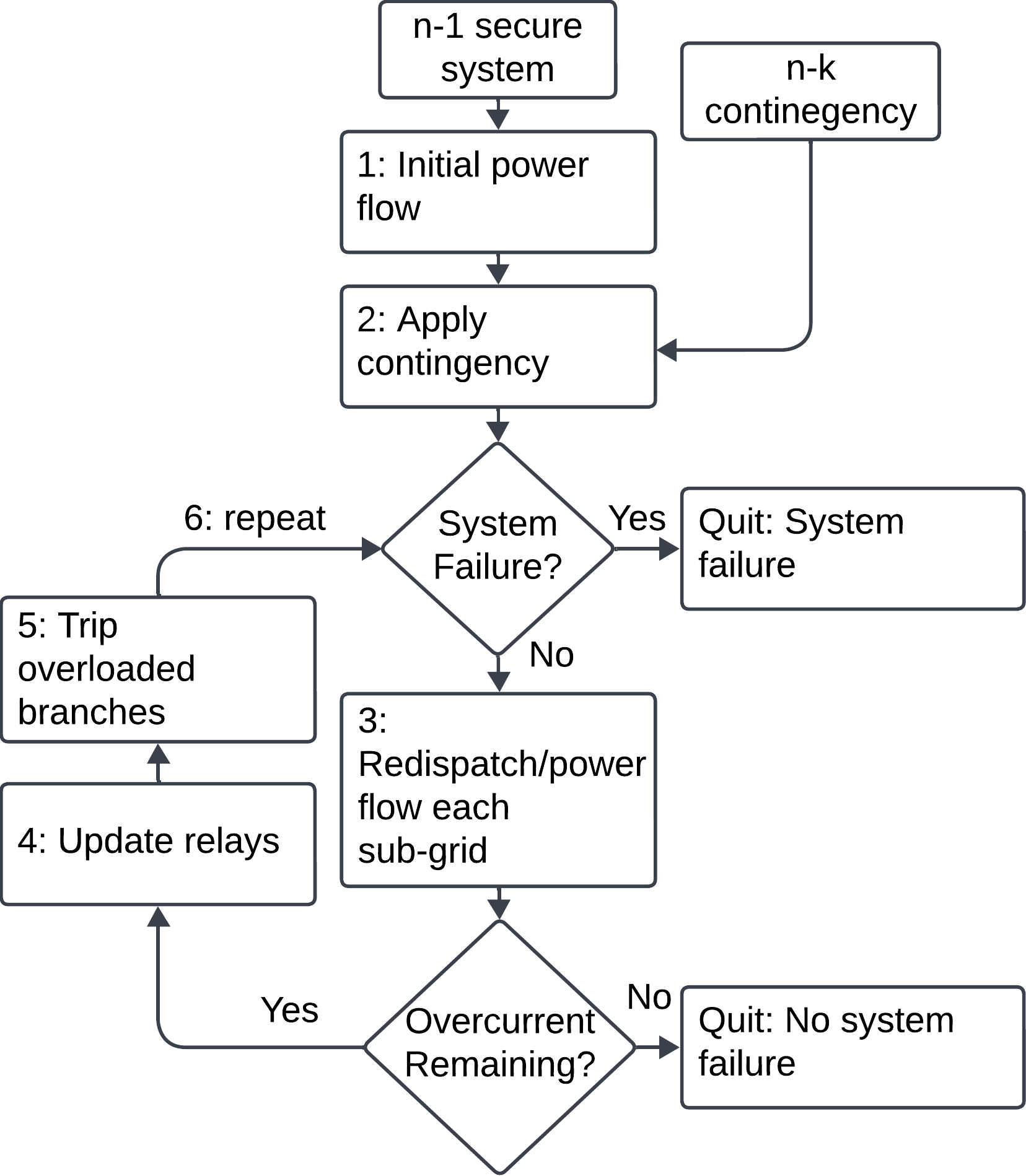}
    \caption{Diagram of the DCIMSEP Cascading Blackout Simulator, adapted from \cite{Eppstein_Hines_2012}.}
    \label{fig:dcsimsep_diagram}
\end{figure}
To test the proposed model variants, we make use of the DCSIMSEP cascading blackout simulator proposed in \cite{Eppstein_Hines_2012} and described in Fig. \ref{fig:dcsimsep_diagram} above. We note that while this simulator makes use of DC power flow, there is no obstacle to the usage of a simulator that includes AC power flow, power system dynamics, or other aspects of power system operations (beyond additional computational complexity). For information on proper benchmarking and validation of cascading simulators, we refer the reader to \cite{Bialek_2016}).

As input to the blackout simulator, we make use of the RTS-GMLC test system \cite{rts-gmlc}. This 73 bus, 120 line test case provides a year's worth of realistic hourly load and generation data mean to mimic a high-renewable-penetration region in the Southwestern United States. 

Our final dataset is constructed by running the simulator for every possible N-2 line-failure contingency (7140 possibilities), and every hourly RTS-GMLC load/generation profile (8784 profiles corresponding to each hour of 2020). Together this yields a total dataset size of 62,717,760 samples. We then perform a 70/15/15 percent train-validate-test split. The data contained in each sample is as follows:
\begin{itemize}
    \item Bus features: Load, Generation
    \item Line features: Resistance, Reactance, Initial Failure State (binary)
    \item Target variable: Blackout Size (MW)
\end{itemize}
Encoding of the input features of differs between the GNN-based regression models and the XGBoost classifier model. In the former, bus and line features are encoded directly in the input graph as node and edge features respectively. In the latter, all features are simply presented as a 1-dimensional vector, with binary initial line failures encoded in a 'multi-hot' fashion.

\subsection{Dataset Analysis}
\begin{figure}
    \centering
    \includegraphics[width=0.9\columnwidth]{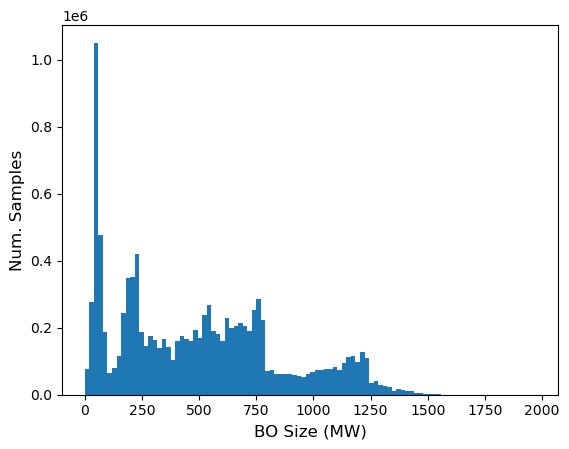}
    \caption{Histogram of Blackout Sizes (Zeroes Excluded)}
    \label{fig:bo-histogram}
\end{figure}
Of the ~62.7M samples in our full dataset, 82.4\% have no blackout. The remaining samples have blackout sizes distributed according to Fig. \ref{fig:bo-histogram}. The mean and median non-zero blackout sizes are 494.1MW and 464.7MW respectively.

\subsection{Calculation and Selection of Statistical Influence Edges}
\label{sec:data-stat}

We sample 6.2 million system states, 
corresponding to all N-2 contingencies for 88 randomly selected load/generation profiles (approximately $\frac{1}{100}$ of the number of system states considered in the full dataset). For these samples, we use the DCIMSEP and generate the same features as above, but in addition keep track of which lines fail both initially and throughout the cascade. We then employ the method described in Section IV to generate a list of candidate statistical edges. As the optimal number of statistical edges to add is initially unknown, we create topologies that contain all physical edges in the system, in addition to 0, 5, 10, and 20 statistical edges.
For illustration, Fig. \ref{fig:rts-statlines} displays the topology of the RTS-GMLC system when augmented with 5 statistical edges. 

\subsection{Training and Tuning: GNN}
\label{sec:tune_GNN}

\begin{table}[]
\caption{GNN Hyper-parameter Grid-Search}
\label{tab:GNN_hp_grid_search}
\centering
\begin{tabular}{|l|l|}
\hline
\textbf{Hyper-Parameter} & \textbf{Tested Values} \\ \hline
\hline
\textbf{Layer Count} & 2, 3, \textbf{4} \\ \hline
\textbf{Learning Rate} & \textbf{0.001}, 0.01, 0.1 \\ \hline
\textbf{Batch Size} & 32, 64, \textbf{128} \\ \hline
\end{tabular}%
\end{table}
We perform a hyper-parameter grid search to determine optimal architecture and training settings for the GNN models, summarized in Table \ref{tab:GNN_hp_grid_search} with selected values given in bold. To enable tuning given the large dataset size, we perform a relatively limited search in two stages. First, we train each of the 27 possible configuration for 2 epochs and evaluate their performance on the validation set. We then select the 3 options with the lowest validation error and train for 5 epochs each, making final hyper-parameter setting decisions based on the results of this second stage. Hyper-parameter tuning was performed separately for GNNs trained on mixed (blackout and no-blackout) and blackout-only versions of the dataset. In both cases, the models configured with 4 graph-convolution layers, a learning rate of 0.001, and a batch size of 128 were determined to be optimal.

We also perform short training runs of 5 epochs for both the mixed GNN and blackout-only GNN to determine the optimal number of statistical edges to include for the augmented-topology versions of each model variant (R+, CR+, and CVR+). In the case of the augmented-topology mixed GNN (to be used in the R+ model, and in the verification step of the CVR+ model), the augmented topology with 10 statistical edges was found to have the lowest validation error (as compared to those with 5 or 20 statistical lines). The augmented-topology blackout-only GNN (to be used in the regression steps of the CR+ and CVR+ models) instead yielded the highest performance when the physical topology was augmented with only 5 statistical lines. As a result, final training runs (on the combined train/validation set) were performed with 10 statistical lines for the augmented-topology mixed GNN, and with 5 statistical lines for the augmented-topology blackout-only GNN model. Standard-topology versions of each (using only the physical topology) were also trained for both the mixed and blackout-only GNNs.

\subsection{Training and Tuning: Classifier}
\label{sec:clf_tune}
\begin{table}[]
\caption{XGBoost Hyper-parameter Grid-Search Results}
\label{tab:xgb_hp_search_val}
\centering
\begin{tabular}{|l|l|}
\hline
\textbf{Hyper-Parameter} & \textbf{Tested Values} \\ \hline
\hline
\textbf{max\_depth} & 2, 4, 6, \textbf{8} \\ \hline
\textbf{min\_child\_weight} &  0, \textbf{1}, 10, 50, 100 \\ \hline
\textbf{gamma} & 0, \textbf{1}, 10, 50, 100\\ \hline
\textbf{sample\_weight} & None, \textbf{Linear} \\ \hline
\end{tabular}%
\end{table}

To determine the optimal configuration for the XGBoost classifier, we perform a grid search over a developer-recommended \cite{XGB_docs} subset of XGBoost hyper-parameters: 'max\_depth', which controls the maximum depth of generated decision trees, 'min\_child\_weight', which sets the minimum total instance weight, and 'gamma', which sets the minimum loss reductions necessary to further split a leaf node. Additionally, though the resultant model will perform only binary classification,  we test the effect of applying linear sample weighting based on blackout size during training. 

Configurations are trained on the train set and evaluated in terms of F1 score on the validation set. Table \ref{tab:xgb_hp_search_val} depicts the tested values, with the optimal configuration displayed in bold.

\begin{table}[]
\setlength{\tabcolsep}{2pt}
\renewcommand{\arraystretch}{1.1}
\caption{Blackout Size Estimation Results for both standard and augmented topologies.}
\label{tab:regression_results_vartop}
\resizebox{\columnwidth}{!}{%
\begin{tabular}{|l|ll|ll|ll|}
\hline
 \textbf{Model} & \multicolumn{2}{l|}{\textbf{\begin{tabular}[c]{@{}l@{}}Sample Error \\ (all samples)\end{tabular}}} & \multicolumn{2}{l|}{\textbf{\begin{tabular}[c]{@{}l@{}}Sample Error \\ (BO samples)\end{tabular}}} & \multicolumn{2}{l|}{\textbf{\begin{tabular}[c]{@{}l@{}}Sample Error \\ (non-BO samples)\end{tabular}}} \\ \hline
 & \multicolumn{1}{l|}{MAE} & MedAE & \multicolumn{1}{l|}{MAE} & MedAE & \multicolumn{1}{l|}{MAE} & MedAE \\ \hline
 & \multicolumn{1}{l|}{} &  & \multicolumn{1}{l|}{} &  & \multicolumn{1}{l|}{} &  \\ \hline
\textbf{R}& \multicolumn{1}{l|}{9.28} & 0.099 & \multicolumn{1}{l|}{40.33} & 17.83 & \multicolumn{1}{l|}{2.51} & 0.099 \\ \hline
\textbf{R+10} & \multicolumn{1}{l|}{8.84} & 0.069 & \multicolumn{1}{l|}{39.33} & 16.67 & \multicolumn{1}{l|}{2.36} & 0.069 \\ \hline
 & \multicolumn{1}{l|}{} &  & \multicolumn{1}{l|}{} &  & \multicolumn{1}{l|}{} &  \\ \hline
\textbf{CR (Perfect Classifier)}& \multicolumn{1}{l|}{} &  & \multicolumn{1}{l|}{} &  & \multicolumn{1}{l|}{} &  \\ \hline
\textbf{CR} & \multicolumn{1}{l|}{3.32} & 0.00 & \multicolumn{1}{l|}{18.91} & 8.77 & \multicolumn{1}{l|}{0.00} & 0.00 \\ \hline
\textbf{CR+5} & \multicolumn{1}{l|}{2.99} & 0.00 & \multicolumn{1}{l|}{17.09} & 7.28 & \multicolumn{1}{l|}{0.00} & 0.00 \\ \hline
 & \multicolumn{1}{l|}{} &  & \multicolumn{1}{l|}{} &  & \multicolumn{1}{l|}{} &  \\ \hline
\textbf{CR (XGBoost Classifier)}& \multicolumn{1}{l|}{} &  & \multicolumn{1}{l|}{} &  & \multicolumn{1}{l|}{} &  \\ \hline
\textbf{CR} & \multicolumn{1}{l|}{10.79} & 0.00 & \multicolumn{1}{l|}{41.05} & 9.35 & \multicolumn{1}{l|}{4.35} & 0.00 \\ \hline
\textbf{CR+5} & \multicolumn{1}{l|}{10.46} & 0.00 & \multicolumn{1}{l|}{39.29} & 7.82 & \multicolumn{1}{l|}{4.33} & 0.00 \\ \hline
 & \multicolumn{1}{l|}{} &  & \multicolumn{1}{l|}{} &  & \multicolumn{1}{l|}{} & \\ \hline
\textbf{CVR}  & \multicolumn{1}{l|}{8.98} & 0.00 & \multicolumn{1}{l|}{21.13} & 8.94 & \multicolumn{1}{l|}{6.40} & 0.00 \\ \hline
\textbf{CVR+5}  & \multicolumn{1}{l|}{8.67} & 0.00 & \multicolumn{1}{l|}{19.35} & 7.44 & \multicolumn{1}{l|}{6.41} & 0.00 \\ \hline
\end{tabular}%
}
\end{table}

\section{Results}
\label{sec:results}
We next examine the test-set performance of the proposed models.
We seek to address the questions that motivated this work. First, we investigate whether statistical augmentation of graph input topology. Second, we assess how incorporation of a pre-regression classification step improves the accuracy of blackout size estimation. 
Further, we analyze two types of undesirable model behavior---severe under-estimations and severe over-estimations of blackout sizes---and discuss how the prevalence and magnitude of these errors varies between our proposed models, as well as the implications that this may have on their fitness for real-world deployment. 

\subsection{Impact of Statistical Topology Augmentation}
First, we analyze model performance across various GNN topologies, as summarized in Table \ref{tab:regression_results_vartop}.
We compare mean absolute error (MAE) and median absolute error (MedAE) for each model variant in different sample categories including all samples, blackout (BO) samples and non-blackout (non-BO) samples. Here we define a blackout sample as one in which a blackout truly occurred. The augmented-topology sub-variants R+, CR+, and CVR+ are labeled with the number of statistical lines contained in their topology, eg R+10 indicates an R model using a topology with 10 statistical lines. For the CR model, we include an assessment both with a ``perfect classifier'' and the realistic XGBoost classifier described above. 

By doing a pairwise comparison of the different models with and without augmented topology, we observe consistent improvements in blackout estimation among models that make use of augmented topologies. Both the MAE and MedAE decrease across all models and all sample categories (all, BO and non-BO samples) with decreases ranging from 3.1\% - 17\% depending on the metric, model and sample category. The only exception is that augmented topology lead to a very small (0.15\%) increase in the MAE on non-blackout samples for the CVR+5 model.

It should be noted that CR model with a perfect classifier has zero error (both MAE and MedAE) in the non-blackout sample category, as these samples are perfectly classified. Furthermore, the MedAE is zero in the ``all samples`` category due to the greater prevalence of non-blackout samples that are correctly classified.
Similarly, MedAE is zero in both the ``all samples`` and ``non-blackout samples`` categories for the CR and CVR models with the XGBoost classifier. However, in this case the zero error results from the performance of the XGBoost Classifier (rather than by assumption as in the perfect-classifier case).

\subsection{Performance of XGBoost Classifier}

Before assessing the performance of our models which employ a classification step (CR and CVR), we briefly examine the performance of the XGBoost classifier separately. The classifier achieves an accuracy of 0.98, indicating that a very high share of the samples are correctly predicted. The classifier precision is 0.96, indicating a false positive rate of 4\% (i.e. 4\% of samples classified as leading to blackout in reality did not). The recall of the classifier is 0.94, indicating that the classifier is able to identify 94\% of samples leadning to blackout. The F1 score is 0.95.

We note that selection of XGBoost hyper-parameters was done via F1 score (the mean of precision and recall) and no attempt was made to target an `optimal' balance of precision and recall. Real-world implementation would likely benefit from more attention in this area. For example, it may be desirable to make adjustments to the hyper-parameters or decision threshold in order to increase recall (i.e. avoid false negatives). Such an adjustment may be worthwhile even at the cost of a decrease in precision, particularly if positive (predicting significant blackout) outputs are verified with an engineering simulation. 

\subsection{Impact of Pre-Regression Classification Step}

\begin{figure}
    \centering
    \begin{subfigure}[t]{\columnwidth}
        \centering
        \includegraphics[width=0.95\columnwidth]{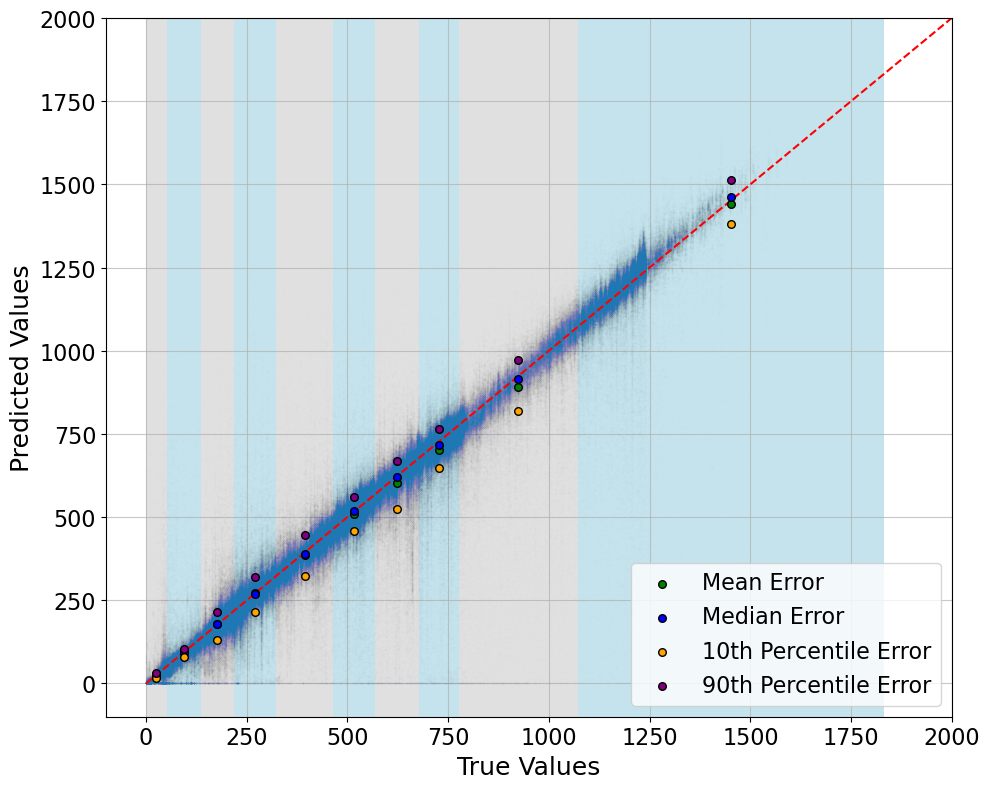} 
        \subcaption{R+ Model with 10 Statistical Edges }
        \label{fig:mixed-parity-plot}
    \end{subfigure}

    \begin{subfigure}[t]{\columnwidth}
        \centering 
        \includegraphics[width=0.95\columnwidth]{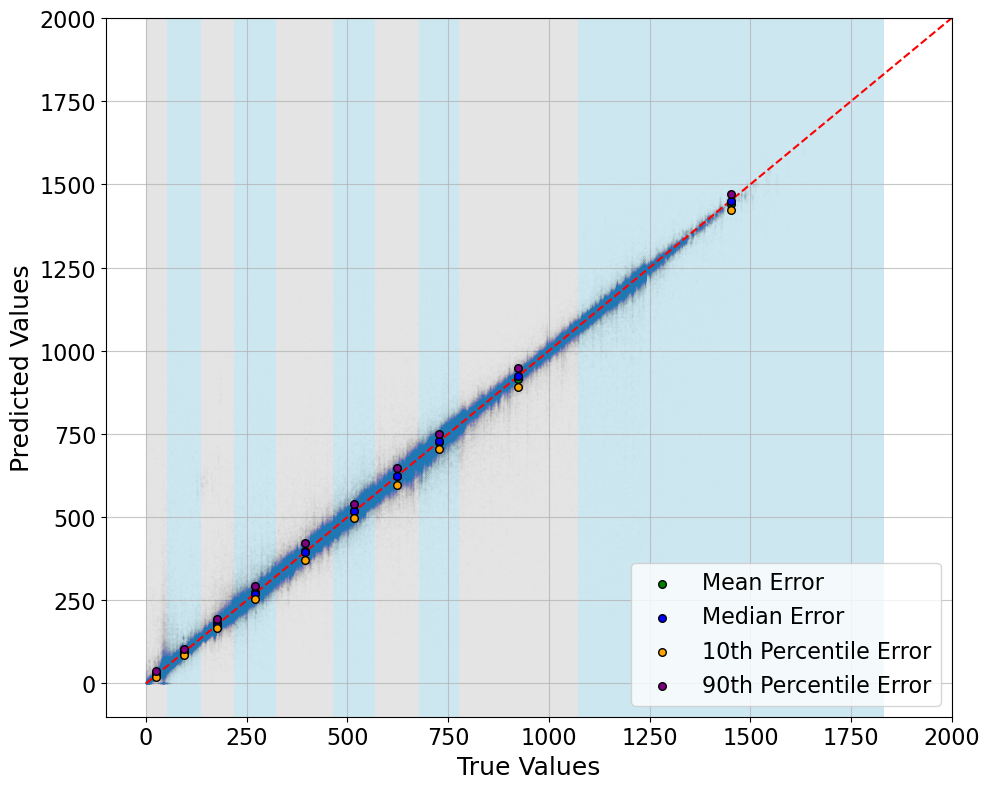}
        \caption{CR+ Model (Perfect Classifier) with 5 Statistical Edges}
        \label{fig:bo-only-parity-plot}
    \end{subfigure}
\caption{\small R+ and CR+ (Perfect Classifier) Parity Plots. The blue dots represent the predicted vs true value for all blackout samples, while the black dots and error bars represent the error statistics across a range of points (indicated by shading in blue or pink).}
\end{figure}

\begin{table}[]
\setlength{\tabcolsep}{2pt}
\renewcommand{\arraystretch}{1.1}
\caption{Blackout Size Estimation Results (Best Topology)}
\label{tab:regression_results_bestop}
\resizebox{\columnwidth}{!}{%
\begin{tabular}{|l|ll|ll|ll|}
\hline
 \textbf{Model}& \multicolumn{2}{l|}{\textbf{\begin{tabular}[c]{@{}l@{}}Sample Error \\ (all samples)\end{tabular}}} & \multicolumn{2}{l|}{\textbf{\begin{tabular}[c]{@{}l@{}}Sample Error \\ (BO samples)\end{tabular}}} & \multicolumn{2}{l|}{\textbf{\begin{tabular}[c]{@{}l@{}}Sample Error \\ (non-BO samples)\end{tabular}}} \\ \hline
 & \multicolumn{1}{l|}{MAE} & MedAE & \multicolumn{1}{l|}{MAE} & MedAE & \multicolumn{1}{l|}{MAE} & MedAE \\ \hline
R+10& \multicolumn{1}{l|}{8.84} & 0.069& \multicolumn{1}{l|}{39.33} & 16.67& \multicolumn{1}{l|}{2.36} & 0.069\\ \hline
CR+5 (Perfect):
& \multicolumn{1}{l|}{2.99} & 0.00 & \multicolumn{1}{l|}{17.09} & 7.28& \multicolumn{1}{l|}{0.00} & 0.00 \\ \hline
CR+5 (XGBoost) 
& \multicolumn{1}{l|}{10.46} & 0.00 & \multicolumn{1}{l|}{39.29} & 7.82& \multicolumn{1}{l|}{4.33} & 0.00 \\ \hline
CVR+5 (XGBoost) & \multicolumn{1}{l|}{8.67} & 0.00 & \multicolumn{1}{l|}{19.35} & 7.44& \multicolumn{1}{l|}{6.41} & 0.00 \\ \hline
 \end{tabular}%
}
\end{table}

To determine the effect of incorporating a classification step, we compare the results of the best-performing GNN input topology for each model type, i.e. R+, CR+ (with perfect classifier), CR+ (with XGBoost classifier), and CVR+. These results are summarized in Table \ref{tab:regression_results_bestop}. 

\subsubsection{With a perfect classifier}
We first compare how the R model (with only regression) compares to the CR model with a perfect classifier and a regression model trained only on blackout samples. The CR model produces drastically lower error rates than the R model, i.e. 2.99MW MAE vs 8.84MW MAE. 
Part of this is explained by the perfect classifier, which effectively assigns a zero error to 82.4\% of our test set (i.e. zero error for all non-blackout samples). 

However, we note that a large improvement was also made among blackout samples, with 
However, we note that the CR model also reduces errors on blackout samples compared with the R model, both for MAE (reduced from 17.09MW to 16.67MW) and MedAE (reduced from 39.33MW to 7.28MW). This indicates that the GNN trained only on blackout samples is better able to estimate the magnitude of load-shed among samples for which load-shed occurs. This pattern can be seen visually across the whole spectrum of blackout sizes in Figures \ref{fig:mixed-parity-plot} and \ref{fig:bo-only-parity-plot}.

\subsubsection{With a realistic classifier}
We next consider the impact of pairing the CR model with the realistic XGBoost Classifier. First, we consider how using the XGBoost impacts performance when predicting the blackout size on non-blackout samples (where the correct size is 0MW). 
As a result of the 96\%  precision rate of the XGBoost classifier, 4\% of negative samples are incorrectly labeled as positive by the classifier. For these samples,
we end up using a GNN trained only on blackout samples to estimate the magnitude of a non-blackout sample. As a result, the MAE of the CR model is 4.33MW for non-blackout samples, compared to 0MW for the CR model with a perfect classifier, and 2.36MW for the R model. This indicates that the blackout-only GNN in the CR models is less capable of correctly identifying no-blackout samples than the mixed GNN of the R model. However, the relatively small difference in performance (4.33MW for CR compared with 2.36MW for R) suggests that 
the blackout-only GNN is still broadly able produce small outputs when exposed to non-blackout samples. 

Next, we consider how using the XGBoost impacts performance when predicting the blackout size on blackout samples (where the correct answer is $>$0MW).
The 94\% recall rate of the XGBoost classifier means that 6\% of positive samples are incorrectly labeled as negative (no blackout). In this case the CR model immediately outputs 0MW, and the sample(s) in question are not fed into the GNN component for blackout magnitude estimation. As a result of this error, the CR model with XGBoost has an MAE of 39.29MW among blackout samples, more than double the blackout-sample MAE of the perfect-classifier CR model (17.09MW) and only slightly lower than that of the R model (39.33MW). Since these errors only affect 6\% of the samples, the median error MedAE of the CR model is still significantly reduced compared with the R model (from 16.67MW to 7.82MW).

\subsubsection{With a realistic classifier and a verification step}
The intention of the CVR model is to reduce the impact of blackout samples falsely labeled as `no blackout' in the initial classification step. By adding the additional `verification' step, the CVR model achieves MAEs of 8.67MW for all samples, 19.35MW for blackout samples, and 6.41 MW for non-blackout samples. These results make CVR the best-performing practically-implementable model (ie not assuming a perfect classifier) across all samples. On blackout samples, the MAE and MedAE (19.35MW and 7.44MW, respectively) approach those of the perfect-classifier CR model (17.09MW and 7.28MW), supporting the usefulness of a verification step in preventing severe under-estimations arising from classifier false-negatives. 

However, while the CVR model performed best across all samples and on blackout samples specifically, it has the greatest tendency to over-estimate on non-blackout samples (6.41MW MAE, versus 2.36MW and 4.33MW for the R and XGBoost-Classifier CR models respectively). This raises the question of which categories of error (for example, over-estimations of no/small-blackout samples versus under-estimations of blackout samples) should be considered most important in selecting a `best' model type for real-world use.

\subsection{Error Analysis: Severe Over- and Under-Estimates}

\begin{table}[]
\setlength{\tabcolsep}{1.5pt}
\caption{Statistics on Severe Errors}
\label{tab:severe_errors_table}
\renewcommand{\arraystretch}{1.2}
\resizebox{\columnwidth}{!}{%
\begin{tabular}{|l|lll|lll|}
\hline
\multicolumn{1}{|c|}{\textbf{Model}} & \multicolumn{3}{c|}{\textbf{Severe Under-Estimates}} & \multicolumn{3}{c|}{\textbf{Severe Over-Estimates}} \\ \hline
 & \multicolumn{1}{l|}{Incidence (\%)} & \multicolumn{1}{l|}{MAE} & MedAE & \multicolumn{1}{l|}{Incidence (\%)} & \multicolumn{1}{l|}{MAE} & MedAE \\ \hline
R +10& \multicolumn{1}{l|}{0.39} & \multicolumn{1}{l|}{346.91} & 223.22 & \multicolumn{1}{l|}{1.09} & \multicolumn{1}{l|}{159.43} & 105.69 \\ \hline
CR (XGBoost) +5& \multicolumn{1}{l|}{5.56} & \multicolumn{1}{l|}{466.55} & 422.76 & \multicolumn{1}{l|}{0.79} & \multicolumn{1}{l|}{547.67} & 549.77 \\ \hline
CVR (XGBoost) +5& \multicolumn{1}{l|}{1.07} & \multicolumn{1}{l|}{250.43} & 164.46 & \multicolumn{1}{l|}{1.15} & \multicolumn{1}{l|}{556.26} & 563.46 \\ \hline
\end{tabular}%
}
\end{table}

\begin{figure}
    \centering
    \begin{subfigure}[t]{\columnwidth}
        \centering
        \includegraphics[width=0.9\columnwidth]{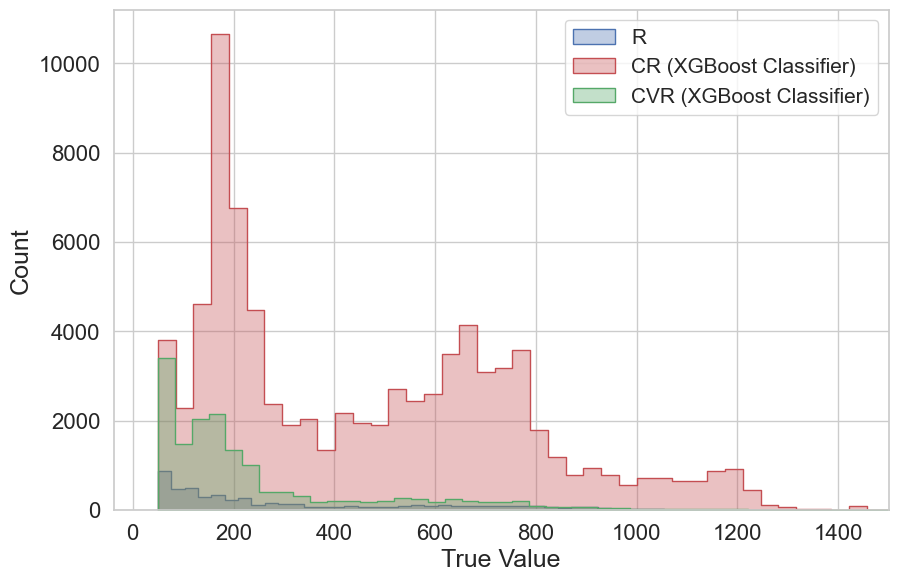} 
        \subcaption{Severe Under-Estimates}
        \label{fig:fn-histogram}
    \end{subfigure}

    \begin{subfigure}[t]{\columnwidth}
        \centering 
        \includegraphics[width=0.9\columnwidth]{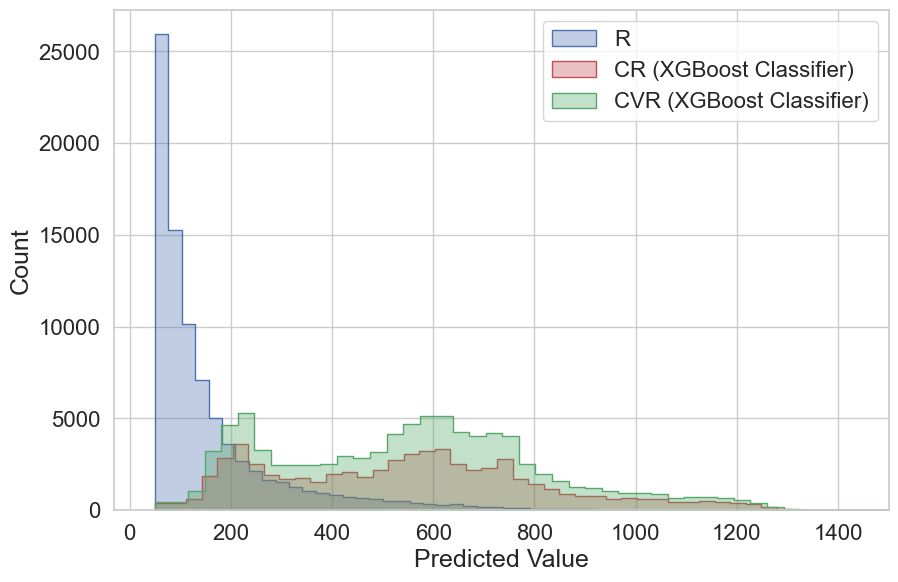}
        \caption{Severe Over-Estimates}
        \label{fig:fp-histogram}
    \end{subfigure}
\caption{Distribution of Severe Errors Across All Model Types}

\end{figure}

We further examine two types of undesireable model behavior: Severe Under-Estimates and Severe Over-Estimates.  For the sake of this analysis, we define Severe Under-Estimates as a predicted value of $<$10MW when the true value is $>$50MW. Similarly, we define Severe Over-Estimates as those for which the true value is $<$10MW and the predicted value is $>$50MW. These threshold values were chosen to divide blackout sizes that may be considered manageable without significant intervention ($<$10MW) from those which may require decisive action ($>$50MW). Defined in this way, severe under/over-estimates can be thought of as the continuous analogue to false negatives/positives.  We note, however, that the thresholds can be chosen differently dependending on the grid in question. 

Table \ref{tab:severe_errors_table} above shows the incidence of severe over- and under-estimates, as well as the error statistics (MAE and MedAE) for the highest-performing versions of each model. We exclude the perfect-classifier version of the CR model as it is not practically implementable. 

In terms of Severe Under-Estimates, the R model shows the lowest incidence rate of 0.39\%, as compared to 5.56\% and 1.07\% for the CR and CVR models respectively. However, the CVR model yields the lowest mean error size of 250.43MW as compared to 356.91MW and 466.55MW for the R and CR models (note that the R model histogram has a heavy tail that is hard to see due to overall small sample count). Figure \ref{fig:fn-histogram} gives histograms of the true values of the samples qualifying as `Severe Under-Estimates' by each model, providing a visual representation of both the comparative prevalence of such errors (by total area), and the distribution of the true blackout sizes incorrectly predicted to be $<$10MW in size. From this, we can clearly see the dramatically higher incidence of Severe Under-Estimates by the CR model as compared to the R and CVR models. By comparing the shapes of the R and CVR true-value distributions, we can see that the CVR model has greater error incidence than the R model, but tends towards lower error magnitudes (as evidenced by its lower mean and median error sizes).

A different pattern emerges when looking at the results for Severe Over-Estimations. The CR model, which had the highest incidence rate for Severe Under-Estimations, now has the lowest: 0.79\% as compared to 1.09\% and 1.15\% for the R and CVR models respectively. The R model comes out significantly ahead in terms of mean error size among Severe Over-Estimations, yielding a mean error value of 159.43MW (versus 547.67MW and 556.26MW for the CR and CVR models respectively). Figure \ref{fig:fp-histogram} depicts the distributions of predicted values among Severe Over-Estimation samples, and illustrates the more comparable incidences of these errors between the models, as well as the tendency of the CR and CVR models to over-estimate blackout sizes by much larger amounts.
The high, and very similar, error values/prediction distributions for the CR and CVR models are likely largely attributable to the 4\% false positive rate of the XGBoost classifier. In the case of such a false positive, a sample with a true value of zero (non-blackout) is input to the (identical) GNN component of the CR/CVR model, which has not been trained on non-blackout samples. 

\subsection{Identifying the best model}
Our results indicate that statistical topology augmentation can reliably increase the ability of GNN-based models to accurately estimate potential blackout size. 

When comparing the augmented-topology models R+, CR+, and CVR+, the results are more mixed. 
The CVR model displayed the lowest overall error, and the greatest ability to correctly estimate the magnitude of samples for which there is a blackout. However, it is the least accurate model on non-blackout samples, as evidenced by a high MAE in this sample group. 
The R model has an overall MAE comparable to the CVR model, and a lower MAE on non-blackout samples. This comes at the expense of significantly higher MAE and MedAE on the blackout samples (i.e. the R model is not able to accurately predict the size of a blackout).

To shed further light on which model-type may be optimal, we consider an operator using our model to screen through a large set of potential contingency scenarios (too large to run engineering simulations for each). If the machine-learning model predicts a blackout size of a concerning level, the contingency scenario is run through a more sophisticated engineering simulation for verification.

Consider first the impact of over-prediction of blackout magnitudes. In this case, we argue that the \emph{incidence} (i.e. number) of errors which trigger an unnecessary simulation is of greater importance than the magnitude of those errors. Using `Severe Over-Estimation' as a proxy for the number of over-predictions, we see all three models have similar incidence rate of such errors (i.e. 0.79\%, 1.09\%, and 1.15\% for the R+, CR+ and CVR+ models). This suggests that an operator could choose to employ the CVR+ model, obtaining the lowest overall errors at the cost of a small increase in unnecessary engineering simulations.

Consider next the impact of large under-estimations of blackout magnitude. In a real-world implementation, large under-estimations are potentially very damaging as a large-blackout scenario may go unnoticed. Similarly to the case of severe under-estimation, it is likely that the incidence rate of such errors is of the greatest concern. From this vantage point the R model appears superior, with a Severe Over-Estimation rate of 0.39\% as compared to 5.56\% for the CR model or 1.07\% for the CVR model. 

In sum, our results indicate that R and CVR models are both likely to provide adequate performance in our example use-case. The choice between them should depend on whether on the user considers overall accuracy (choose CVR) or achieving a minimum severe error rate (choose R) to be most important.

\section{Conclusions}
\label{sec:conclusions}
This paper proposes several new GNN-based models for prediction of potential blackout size from initial grid conditions/failures. 
We propose a new method to generate `statistical lines' which can be used to augment the GNN input topologies, and demonstrate that such topology augmentation consistently increases performance across all proposed model types.

We further examine how
the inclusion of a pre-regression classification step impacts performance. 
We observe that a GNN model trained only on blackout samples can more accurately estimate blackout sizes compared to a GNN-based model trained on both blackout and no-blackout samples. However, coupling this more accurate GNN model with a realistic classifier, we observe that such a `classify, then regress' approach can increase the incidence of large under/over-predictions. To mitigate the impact of incorrectly classified samples, we propose to use the GNN models as a classifier `verification' step, and demonstrate that this reduces the impact of falsely classified samples. 

These promising results suggest a wide variety directions for future work. First, we plan to test the integration of our models into decision-making frameworks, for example, as method for rapidly screening the blackout potential of all N-X contingencies at a current or forecasted operating state. 
We would like to further develop our approach to classification as an initial step to estimating blackout size---for example implementing joint training strategies to synergize the workings of the classifier and regressor components of our models. 
We believe that there is great potential to improve our method of statistical line generation and to examine alternative modeling approaches.

Finally, we will investigate whether the proposed method generalizes to variable initial failure sizes and multiple topologies/test systems.

\bibliographystyle{IEEEtran}
\bibliography{refs}

\begin{thebibliography}{10}
\providecommand{\url}[1]{#1}
\csname url@samestyle\endcsname
\providecommand{\newblock}{\relax}
\providecommand{\bibinfo}[2]{#2}
\providecommand{\BIBentrySTDinterwordspacing}{\spaceskip=0pt\relax}
\providecommand{\BIBentryALTinterwordstretchfactor}{4}
\providecommand{\BIBentryALTinterwordspacing}{\spaceskip=\fontdimen2\font plus
\BIBentryALTinterwordstretchfactor\fontdimen3\font minus
  \fontdimen4\font\relax}
\providecommand{\BIBforeignlanguage}[2]{{%
\expandafter\ifx\csname l@#1\endcsname\relax
\typeout{** WARNING: IEEEtran.bst: No hyphenation pattern has been}%
\typeout{** loaded for the language `#1'. Using the pattern for}%
\typeout{** the default language instead.}%
\else
\language=\csname l@#1\endcsname
\fi
#2}}
\providecommand{\BIBdecl}{\relax}
\BIBdecl

\bibitem{OPA}
I.~Dobson, B.~Carreras, V.~Lynch, and D.~Newman, ``An initial model fo complex
  dynamics in electric power system blackouts,'' in \emph{Hawaii International
  Conference on System Sciences}, 2001, pp. 710--718.

\bibitem{Eppstein_Hines_2012}
M.~J. Eppstein and P.~D.~H. Hines, ``A “random chemistry” algorithm for
  identifying collections of multiple contingencies that initiate cascading
  failure,'' \emph{IEEE Transactions on Power Systems}, vol.~27, no.~3, p.
  1698–1705, Aug 2012.

\bibitem{Kirschen_2004}
D.~Kirschen, D.~Jayaweera, D.~Nedic, and R.~Allan, ``A probabilistic indicator
  of system stress,'' \emph{IEEE Transactions on Power Systems}, vol.~19,
  no.~3, pp. 1650--1657, 2004.

\bibitem{Chen_Mili_2013}
Q.~Chen and L.~Mili, ``Composite power system vulnerability evaluation to
  cascading failures using importance sampling and antithetic variates,''
  \emph{IEEE Transactions on Power Systems}, vol.~28, no.~3, p. 2321–2330,
  Aug 2013.

\bibitem{Henneaux_Labeau_2014}
P.~Henneaux and P.-E. Labeau, ``Improving monte carlo simulation efficiency of
  level-i blackout probabilistic risk assessment,'' in \emph{2014 International
  Conference on Probabilistic Methods Applied to Power Systems (PMAPS)}, Jul
  2014, p. 1–6.

\bibitem{Kim_Bucklew_Dobson_2013}
J.~Kim, J.~A. Bucklew, and I.~Dobson, ``Splitting method for speedy simulation
  of cascading blackouts,'' in \emph{2013 IEEE Power \& Energy Society General
  Meeting}, Jul 2013, p. 1–1.

\bibitem{Wang_Chen_Liu_Chen_Shortle_Wu_2015}
S.-P. Wang, A.~Chen, C.-W. Liu, C.-H. Chen, J.~Shortle, and J.-Y. Wu,
  ``Efficient splitting simulation for blackout analysis,'' \emph{IEEE
  Transactions on Power Systems}, vol.~30, no.~4, p. 1775–1783, Jul 2015.

\bibitem{Rezaei_Hines_Eppstein_2015}
P.~Rezaei, P.~D.~H. Hines, and M.~J. Eppstein, ``Estimating cascading failure
  risk with random chemistry,'' \emph{IEEE Transactions on Power Systems},
  vol.~30, no.~5, p. 2726–2735, Sep 2015.

\bibitem{Dobson2003}
I.~Dobson, B.~Carreras, and D.~Newman, ``A probabilistic loading-dependent
  model of cascading failure and possible implications for blackouts,'' in
  \emph{36th Annual Hawaii International Conference on System Sciences, 2003.
  Proceedings of the}, Jan 2003, pp. 10 pp.--.

\bibitem{Dobson_Carreras_Newman_2004}
I.~Dobson, B.~A. Carreras, and D.~E. Newman, ``A branching process
  approximation to cascading load-dependent system failure,'' in
  \emph{Proceedings of the 37th Annual Hawaii International Conference on
  System Sciences}, jan 2004, pp. 1--10.

\bibitem{Xiao_Hongda_Yeh_2011}
H.~Xiao and E.~M. Yeh, ``Cascading link failure in the power grid: A
  percolation-based analysis,'' in \emph{2011 IEEE International Conference on
  Communications Workshops (ICC)}, 2011, pp. 1--6.

\bibitem{Kong_Yeh_2012}
Z.~Kong and E.~M. Yeh, ``Correlated and cascading node failures in random
  geometric networks: A percolation view,'' in \emph{2012 Fourth International
  Conference on Ubiquitous and Future Networks (ICUFN)}, Jul 2012, p.
  520–525.

\bibitem{Wang_Zhou_Hu_2018}
Z.~Wang, D.~Zhou, and Y.~Hu, ``Group percolation in interdependent networks,''
  \emph{Physical Review E}, vol.~97, no.~3, p. 032306, Mar 2018.

\bibitem{Zhang_Yeh_Modiano_2019}
J.~Zhang, E.~Yeh, and E.~Modiano, ``Robustness of interdependent random
  geometric networks,'' \emph{IEEE Transactions on Network Science and
  Engineering}, vol.~6, no.~3, p. 474–487, Jul 2019.

\bibitem{hines_cascading_2017}
P.~Hines, I.~Dobson, and P.~Rezaei, ``Cascading power outages propagate locally
  in an influence graph that is not the actual grid topology,'' in \emph{2017
  {IEEE} {Power} \& {Energy} {Society} {General} {Meeting}}, Jul. 2017, pp.
  1--1.

\bibitem{Zhou_Kai_Dobson_2020}
K.~Zhou, I.~Dobson, Z.~Wang, A.~Roitershtein, and A.~P. Ghosh, ``A markovian
  influence graph formed from utility line outage data to mitigate large
  cascades,'' \emph{IEEE Transactions on Power Systems}, vol.~35, no.~4, pp.
  3224--3235, 2020.

\bibitem{Qi_Junjian_2015}
J.~Qi, K.~Sun, and S.~Mei, ``An interaction model for simulation and mitigation
  of cascading failures,'' \emph{IEEE Transactions on Power Systems}, vol.~30,
  no.~2, pp. 804--819, 2015.

\bibitem{Ju_Qi_Junjian_2015}
W.~Ju, J.~Qi, and K.~Sun, ``Simulation and analysis of cascading failures on an
  npcc power system test bed,'' in \emph{2015 IEEE Power \& Energy Society
  General Meeting}, 2015, pp. 1--5.

\bibitem{Liu_Zhang_Wu_Botterud_Yao_Kang_2021}
Y.~Liu, N.~Zhang, D.~Wu, A.~Botterud, R.~Yao, and C.~Kang, ``Searching for
  critical power system cascading failures with graph convolutional network,''
  \emph{IEEE Transactions on Control of Network Systems}, vol.~8, no.~3, p.
  1304–1313, Sep 2021.

\bibitem{Zhu_Zhou_Wei_Wang_2023}
Y.~Zhu, Y.~Zhou, W.~Wei, and N.~Wang, ``Cascading failure analysis based on a
  physics-informed graph neural network,'' \emph{IEEE Transactions on Power
  Systems}, vol.~38, no.~4, p. 3632–3641, Jul 2023.

\bibitem{Shuvro_Das_Hayat_Talukder_2019}
R.~A. Shuvro, P.~Das, M.~M. Hayat, and M.~Talukder, ``Predicting cascading
  failures in power grids using machine learning algorithms,'' in \emph{2019
  North American Power Symposium (NAPS)}, Oct 2019, p. 1–6.

\bibitem{Varbella_Gjorgiev_Sansavini_2023}
A.~Varbella, B.~Gjorgiev, and G.~Sansavini, ``\BIBforeignlanguage{en}{Geometric
  deep learning for online prediction of cascading failures in power grids},''
  \emph{\BIBforeignlanguage{en}{Reliability Engineering \& System Safety}},
  vol. 237, p. 109341, Sep 2023.

\bibitem{Zhu_Zhou_Wei_Zhang_2022}
Y.~Zhu, Y.~Zhou, W.~Wei, and L.~Zhang, ``Real-time cascading failure risk
  evaluation with high penetration of renewable energy based on a graph
  convolutional network,'' \emph{IEEE Transactions on Power Systems}, p.
  1–12, 2022.

\bibitem{Ahmad_Papadopoulos_2022}
T.~Ahmad and P.~N. Papadopoulos, ``Prediction of cascading failures and
  simultaneous learning of functional connectivity in power system,'' in
  \emph{2022 IEEE PES Innovative Smart Grid Technologies Conference Europe
  (ISGT-Europe)}, Oct 2022, p. 1–5.

\bibitem{Chen_Thorp_Parashar_2001}
J.~Chen, J.~Thorp, and M.~Parashar, ``Analysis of electric power system
  disturbance data,'' in \emph{Proceedings of the 34th Annual Hawaii
  International Conference on System Sciences}, Jan 2001, p. 738–744.

\bibitem{Carreras_Newman_Dobson_Poole_2001}
B.~Carreras, D.~Newman, I.~Dobson, and A.~Poole, ``Evidence for self-organized
  criticality in electric power system blackouts,'' in \emph{Proceedings of the
  34th Annual Hawaii International Conference on System Sciences}, Jan 2001, p.
  705–709.

\bibitem{Talukdar_Apt_Ilic_Lave_Morgan_2003}
S.~N. Talukdar, J.~Apt, M.~Ilic, L.~B. Lave, and M.~G. Morgan, ``Cascading
  failures: Survival versus prevention,'' \emph{The Electricity Journal},
  vol.~16, no.~9, p. 25–31, Nov 2003.

\bibitem{MindsUW}
``{Minds@UW},'' \url{https://minds.wisconsin.edu/}, 2024.

\bibitem{kipf_welling_2017}
T.~N. Kipf and M.~Welling, ``Semi-supervised classification with graph
  convolutional networks,'' in \emph{International Conference on Learning
  Representations (ICLR)}, 2017.

\bibitem{velickovic_2018}
P.~Veli{\v{c}}kovi{\'c}, G.~Cucurull, A.~Casanova, A.~Romero, P.~Li{\`o}, and
  Y.~Bengio, ``Graph attention networks,'' in \emph{International Conference on
  Learning Representations}, 2018.

\bibitem{defferrard_2016}
M.~Defferrard, X.~Bresson, and P.~Vandergheynst, ``Convolutional neural
  networks on graphs with fast localized spectral filtering,'' in
  \emph{Advances in Neural Information Processing Systems}, 2016, pp.
  3844--3852.

\bibitem{xu_hu_2019}
K.~Xu, W.~Hu, J.~Leskovec, and S.~Jegelka, ``How powerful are graph neural
  networks?'' in \emph{Proceedings of the International Conference on Learning
  Representations (ICLR)}, 2019.

\bibitem{pfaff_2021}
T.~Pfaff, M.~Fortunato, A.~Sanchez-Gonzalez, and P.~Battaglia, ``Learning
  mesh-based simulation with graph networks,'' in \emph{International
  Conference on Learning Representations}, 2021.

\bibitem{chen_2016}
T.~Chen and C.~Guestrin, ``Xgboost: A scalable tree boosting system,'' in
  \emph{Proceedings of the 22nd ACM SIGKDD International Conference on
  Knowledge Discovery and Data Mining}, 2016, pp. 785--794.

\bibitem{shwartz_armon_2022}
R.~Shwartz-Ziv and A.~Armon, ``Tabular data: Deep learning is not all you
  need,'' \emph{Information Fusion}, vol.~81, pp. 84--90, 2022.

\bibitem{Grinsztajn_2022}
L.~Grinsztajn, E.~Oyallon, and G.~Varoquaux, ``Why do tree-based models still
  outperform deep learning on typical tabular data?'' in \emph{Thirty-sixth
  Conference on Neural Information Processing Systems Datasets and Benchmarks
  Track}, 2022.

\bibitem{Bialek_2016}
J.~Bialek, E.~Ciapessoni, D.~Cirio, E.~Cotilla-Sanchez, C.~Dent, I.~Dobson,
  P.~Henneaux, P.~Hines, J.~Jardim, S.~Miller, M.~Panteli, M.~Papic, A.~Pitto,
  J.~Quiros-Tortos, and D.~Wu, ``Benchmarking and validation of cascading
  failure analysis tools,'' \emph{IEEE Transactions on Power Systems}, vol.~31,
  no.~6, pp. 4887--4900, 2016.

\bibitem{rts-gmlc}
C.~Barrows, A.~Bloom, A.~Ehlen, J.~Ikäheimo, J.~Jorgenson, D.~Krishnamurthy,
  J.~Lau, B.~McBennett, M.~O'Connell, E.~Preston, A.~Staid, G.~Stephen, and
  J.-P. Watson, ``The ieee reliability test system: A proposed 2019 update,''
  \emph{IEEE Transactions on Power Systems}, vol.~PP, pp. 1--1, 07 2019.

\bibitem{XGB_docs}
\BIBentryALTinterwordspacing
D.~D. M.~L. Community, ``Official xgboost documentation.'' [Online]. Available:
  \url{https://xgboost.readthedocs.io/en/stable}
\BIBentrySTDinterwordspacing

\end{thebibliography}

\end{document}